\documentclass[preprint,aps,prd,showpacs,superscriptaddress,nofootinbib,tightenlines]{revtex4}
\usepackage{amsfonts}
\usepackage{multirow}
\usepackage{mathrsfs}
\usepackage{graphicx}
\usepackage{amsmath}
\usepackage{amssymb}
\usepackage{bm}
\usepackage{bbm}
\usepackage{slashbox}

\def\bfsigma{\mbox{\boldmath $\sigma$}}

\def\bfepsilon{\mbox{\boldmath $\epsilon$}}

\def\OMIT#1{}

\newcommand{\nn}{\nonumber}

\newcommand{\beq}{\begin{equation}}
\newcommand{\eeq}{\end{equation}}
\newcommand{\bqa}{\begin{eqnarray}}
\newcommand{\eqa}{\end{eqnarray}}

\begin{document}
%\preprint{}
%%%%%%%%%%%%%%%%%%%%%%%%%%%%%%%%%%%%%%%%%%%%%%%%%%%%%%%%%%%%%%%%%%%%%%%%%%%%%%
\title{\mbox{}\\[10pt]
Reconciling the nonrelativistic QCD prediction and the
$\bm{J}\bm{/}\bm{\psi}\bm{\to}\bm{3}\bm{\gamma}$ data}

%%%%%%%%%%%%%%%%%%%%%%%%%%%%%%%%%%%%%%%%%%%%%%%%%%%%%%%%%%%%%%%%%%%%%%%%%%%%%%

\author{Feng Feng\footnote{fengf@ihep.ac.cn}}
\affiliation{Center for High Energy Physics, Peking University, Beijing 100871, China\vspace{0.2cm}}

\author{Yu Jia\footnote{jiay@ihep.ac.cn}}
\affiliation{Institute of High Energy Physics, Chinese Academy of
Sciences, Beijing 100049, China\vspace{0.2cm}}
\affiliation{Theoretical Physics Center for Science Facilities,
Chinese Academy of Sciences, Beijing 100049, China\vspace{0.2cm}}

\author{Wen-Long Sang\footnote{wlsang@ihep.ac.cn}}
\affiliation{Institute of High Energy Physics, Chinese Academy of
Sciences, Beijing 100049, China\vspace{0.2cm}}
\affiliation{Theoretical Physics Center for Science Facilities,
Chinese Academy of Sciences, Beijing 100049, China\vspace{0.2cm}}

\date{\today}
%%%%%%%%%%%%%%%%%%%%%%%%%%%%%%%%%%%%%%%%%%%%%%%%%%%%%%%%%%%%%%%%%%%%%%%%%%%%%%
\begin{abstract}
It has been a long-standing problem that the rare electromagnetic
decay process $J/\psi\to 3\gamma$ is plagued with both large and
negative radiative and relativistic corrections. To date it remains
futile to make a definite prediction to confront with the branching
fraction of $J/\psi\to 3\gamma$ recently measured by the
\textsf{CLEO-c} and \textsf{BESIII} Collaborations. In this work, we
investigate the joint perturbative and relativistic correction ({\it
i.e.} the ${\mathcal O}(\alpha_s v^2)$ correction, where $v$ denotes
the characteristic velocity of the charm quark inside the $J/\psi$)
for this decay process, which turns out to be very significant.
After incorporating the contribution from this new ingredient, with
the reasonable choice of the input parameters, we are able to
account for the measured decay rates in a satisfactory degree.
\end{abstract}

%%%%%%%%%%%%%%%%%%%%%%%%%%%%%%%%%%%%%%%%%%%%%%%%%%%%%%%%%%%%%%%%%%%%%%%%%%%%%%

%11.10.Hi Renormalization group evolution of parameters
%12.38.Bx Perturbative calculations
%12.38.Cy Summation of perturbation theory
%12.38.-t Quantum chromodynamics
%12.39.St Factorization
%13.20.Gd Decays of J/¦×, ¦´, and other quarkonia
%13.40.Hq Electromagnetic decays
%13.60.Le Meson production
%13.60.Hb Total and inclusive cross sections (including deep-inelastic processes)
%13.87.Fh Fragmentation into hadrons
%14.40.Pq Heavy quarkonia
%%%%%%%%%%%%%%%%%%%%%%%%%%%%%%%%%%%%%%%%%%%%%%%%%%%%%%%%%%%%%%%%%%%%%%%%%%%%%%

\maketitle

Calculating the ortho-positronium (o-Ps) annihilation decay into
three photons, together with the hyperfine splitting of the
positronium, has played a pivotal role in developing the bound-state
QED formalism, particularly in establishing the modern theoretical
framework such as nonrelativisitc QED (NRQED)~\cite{Caswell:1985ui}.
After incorporating a number of higher order
corrections~\cite{Caswell:1976nx,Adkins:2005eg,Adkins:2000fg}, the
theoretical prediction to the decay rate for o-Ps$\to 3\gamma$ is
consistent with the measurement~\cite{Jinnouchi:2003hr} to an
impressive accuracy, which can be viewed as a great triumph of
NRQED.

The QCD analogue of the o-Ps, the $J/\psi$ meson, has also occupied
a special stage in the making of the Standard Model. The numerous
decay channels of the $J/\psi$ have been intensively studied in the
past four decades, among which the electromagnetic decay $J/\psi\to
3\gamma$ is of special interest. Despite considerable efforts,
experimentalists have not observed this rare decay channel until
quite recently. In 2008, the \textsf{CLEO-c} collaboration measured
the branching fraction of this process for the first
time~\cite{Adams:2008aa} and obtained ${\rm Br}(J/\psi\to 3\gamma)=
(1.2\pm 0.3\pm 0.2)\times 10^{-5}$. With much larger $J/\psi$
samples that come from the pionic transition of $\psi'$,
\textsf{BESIII} Collaboration recently refined the earlier
measurement and gave ${\rm Br}(J/\psi\to 3\gamma)=(11.3\pm 1.8\pm
2.0)\times 10^{-6}$~\cite{Ablikim:2012sn}. Both experiments are
compatible with each other, nailing down the branching fraction for
this rare decay to be about $10^{-5}$.

 The theoretical
investigation on $J/\psi\to 3\gamma$ preceded the experimental
discovery long earlier. By the early 1980s, both the first-order
radiative and relative corrections have already been calculated in
the context of the potential model~\cite{Mackenzie:1981sf,
Keung:1982jb,Kwong:1987ak}.
%-----------------------------
%-----------------------------
Just analogous to the very successful application of NRQED to
positronium, nowadays the nonrelativistic QCD (NRQCD) effective
field theory~\cite{Caswell:1985ui} becomes the standard tool to
analyze the quarkonium spectrum, decay and
production~\cite{Brambilla:2010cs}. In particular, the influential
NRQCD factorization approach~\cite{Bodwin:1994jh} has superseded the
phenomenological potential model as a modern and systematic
framework. To the best of our knowledge, up to the error of relative
order $v^4$, the decay rate of the process $J/\psi\to 3\gamma$ in
the NRQCD formalism can be written as
%-----------------------------
\bqa
%-----------------------------
\Gamma(J/\psi\to 3\gamma) &=& {8(\pi^2-9)e_c^6\alpha^3\over 9 m_c^2}
\left|\langle 0| \chi^\dagger \bfsigma \cdot
\bfepsilon^\ast\psi|J/\psi(\bfepsilon)\rangle \right|^2 \Bigg\{1-
12.630\,{\alpha_s\over \pi}
%-----------------------------
\nn\\
%-----------------------------
&&+ \bigg[ {132-19\pi^2 \over 12 (\pi^2-9)} + \bigg( {16 \over 9
}\ln {\mu_f^2 \over m_c^2}+ G \bigg) \,{\alpha_s \over \pi}\bigg]
\langle v^2\rangle_{J/\psi} +\cdots \Bigg\},
%-----------------------------
\label{NRQCD:factorization:decay:rate}
%-----------------------------
\eqa
%-----------------------------
where $e_c={2\over 3}$ and $m_c$ represent the electric charge and
the mass of the charm quark. $\langle 0| \chi^\dagger\bfsigma\cdot
\bm{\epsilon}^\ast\psi|J/\psi(\bfepsilon)\rangle$ is the lowest
order (LO) NRQCD $J/\psi$-to-vacuum matrix element, where
$\bfepsilon$ represents the polarization vector of $J/\psi$. Up to
an error of relative order $v^2$, this matrix element can be
approximated by $\sqrt{3\over 2\pi} R_{J/\psi}(0)$, where
$R_{J/\psi}(0)$ denotes the wave function at the origin for the
$J/\psi$ in the potential model. A useful ingredient of this NRQCD
formula is the inclusion of the leading relativistic correction,
characterized by the ratio of the following NRQCD matrix elements:
%-----------------------------
\bqa
%-----------------------------
\langle v^2\rangle_{J/\psi} & \equiv & {\langle 0| \chi^\dagger
\bfsigma\cdot \bfepsilon^\ast
\big(-\tfrac{i}{2}\overleftrightarrow{\bf{D}}\big)^2 \psi|J/\psi
(\bfepsilon)\rangle \over m_c^2\langle 0| \chi^\dagger \bfsigma\cdot
\bfepsilon^\ast\psi|J/\psi(\bfepsilon)\rangle},
%-----------------------------
\label{v2:definition}
%-----------------------------
\eqa
%-----------------------------
where $\overleftrightarrow{\bf{D}}$ stands for the left-right
asymmetric covariant derivative. The Gremm-Kapustin
relation~\cite{Gremm:1997dq}, which stems from the NRQCD equation of
motion, can be employed to express $\langle v^2\rangle_{J/\psi}$ in
terms of the $J/\psi$ mass and the charm quark pole mass.
Empirically, one may expect that $\langle v^2\rangle_{J/\psi}\approx
0.3$, compatible with the typical velocity of charmonium deduced
from the potential model.

The leading term in (\ref{NRQCD:factorization:decay:rate}) is
translated from the analogous formula for the ortho-positronium
decay to 3 photons, originally derived by Ore and Powell in
1949~\cite{Ore:1949te}. The ${\cal O}(\alpha_s)$
correction~\cite{Kwong:1987ak} was adapted from the ${\cal
O}(\alpha)$ correction to o-Ps$\to 3\gamma$, first correctly
calculated by Caswell, Lepage and Sapirstein in
1977~\cite{Caswell:1976nx}, or from the ${\cal O}(\alpha_s)$
correction to $J/\psi \to 3g$, first calculated by Mackenzie and
Lepage in 1981~\cite{Mackenzie:1981sf}. Note at that time the ${\cal
O}(\alpha_s)$ short-distance coefficient can be obtained only with
very limited precision.
%--------------------------------------------
The relative order-$v^2$ correction was originally obtained by Keung
and Muzinich in 1982~\cite{Keung:1982jb}. With the aid of the
Gremm-Kapustin relation, their result is consistent with the value
given in (\ref{NRQCD:factorization:decay:rate}), ${132-19\pi^2 \over
12 (\pi^2-9)}\langle v^2\rangle_{J/\psi}\approx -5.32 \langle
v^2\rangle_{J/\psi}$~\cite{Schuler:1994hy}.

%--------------------------
\begin{figure}[t]
\begin{center}
\includegraphics[height= 8 cm]{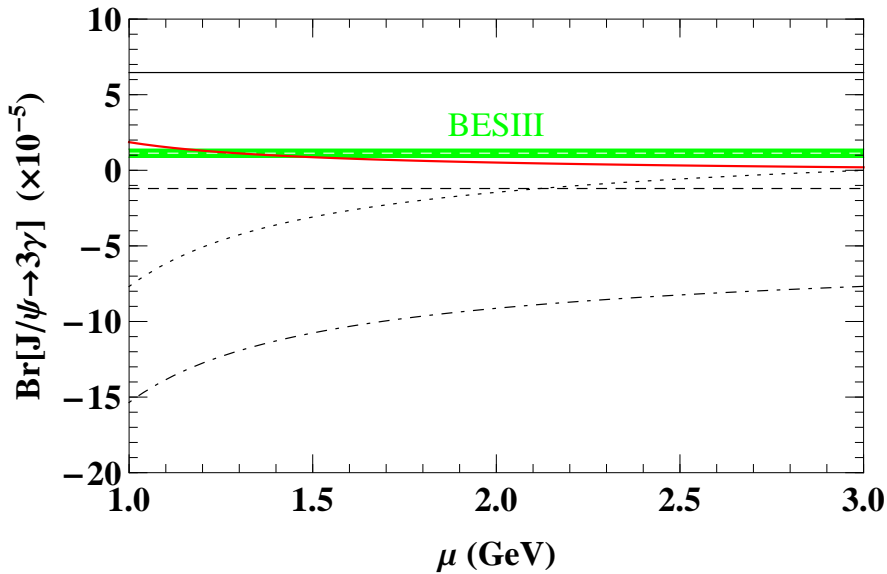}
\caption{The renormalization scale dependence of the branching
fraction for $J/\psi\to 3\gamma$. The LO prediction is represented
by the black solid line, the LO result plus the ${\cal O}(v^2)$
correction by the dashed line, the LO result plus the ${\cal
O}(\alpha_s)$ correction by the dotted curve, the prediction
incorporating both the ${\cal O}(\alpha_s)$ and ${\cal O}(v^2)$
corrections by the dot-dashed curve. Finally, the complete NRQCD
prediction incorporating the new ${\cal O}(\alpha_s v^2)$ correction
is represented by the red solid curve. For the sake of comparison,
the \textsf{BESIII} measurement of the decay branching fraction is
represented by the green band, with statistical and systematic
errors added in quadrature. \label{Various:predictions:depend:mu}}
\end{center}
\end{figure}
%--------------------------

Our current understanding of $J/\psi\to 3\gamma$ is essentially not
much better than three decades ago, since to date the ${\cal
O}(\alpha_s)$ and ${\cal O}(v^2)$ corrections remain to be the only
known corrections. Upon confronting
(\ref{NRQCD:factorization:decay:rate}) with the $J/\psi\to 3\gamma$
data, the agreement seems far less satisfactory than that for its
ortho-positronium cousin. The situation is most clearly illustrated
in Fig.~\ref{Various:predictions:depend:mu}. With some reasonable
choices of the NRQCD matrix elements, the LO prediction is several
times greater than the new \textsf{BESIII} measurement. After
including the ${\mathcal O}(\alpha_s)$ correction, the decay rate
becomes significantly lower than the measured one, and even turns
negative for most plausible range of the renormalization scale
entering the $\alpha_s$. The situation deteriorates desperately
after the ${\cal O}(v^2)$ correction is further included, in which
for virtually all the plausible input parameters, the decay rate
becomes deeply negative thus loses physical significance.

The utter failure of predicting the decay rate for $J/\psi\to
3\gamma$ may imply the breakdown of the NRQCD for this decay
process. The symptom seems to be rooted in the uncomfortably large
negative radiative and relativistic corrections. In 1997, Braaten
and Chen estimated the asymptotic behavior of the higher order
radiative corrections to the coefficient associated with the LO
NRQCD matrix element, and found the series are badly divergent due
to the large residue of the $u=-{1\over 2}$ infrared
renormalon~\cite{Braaten:1998au}.

Since the large negative relativistic correction appears to be even
more troublesome than the radiative correction, in this work we
attempt to investigate the ${\cal O}(\alpha_s v^2)$ effect, by
computing the first-order radiative correction to the short-distance
coefficient associated with $\langle v^2\rangle_{J/\psi}$. Thus far,
the ${\cal O}(\alpha_s v^2)$ correction has only been available for
a few reactions involving $S$-wave quarkonium, {\it e.g.},
$J/\psi\to e^+e^-$~\cite{Luke:1997ys,Bodwin:2008vp}, $B_c \to
l\bar{\nu}$~\cite{Lee:2010ts}, $\eta_c\to
\gamma\gamma$~\cite{Jia:2011ah,Guo:2011tz}, $\eta_c\to$ light
hadrons~\cite{Guo:2011tz,Li:2012rn}, $e^+e^- \to
J/\psi+\eta_c$~\cite{Dong:2012xx}. The ${\cal O}(\alpha_s v^2)$
correction turns out to be insignificant for most of the
aforementioned processes. Nevertheless, for the hadronic decay of
$\eta_c$, the ${\cal O}(\alpha_s v^2)$ correction is found to be
sizable~\cite{Guo:2011tz}. Therefore, it is also of some interest to
assess the impact of the ${\cal O}(\alpha_s v^2)$ correction for
$J/\psi\to 3\gamma$.

The ${\cal O}(\alpha_s v^2)$ coefficient is represented by the
entity in the parenthesis in (\ref{NRQCD:factorization:decay:rate}).
The first term that depends on the NRQCD factorization scale
$\mu_f$, ${16 \over 9 }\ln {\mu_f^2 \over m_c^2}$, can be inferred
solely based on the knowledge of the one-loop UV divergence of the
lowest-order NRQCD matrix element~\cite{Luke:1997ys} and the
Ore-Powell result. Therefore, our central task is to calculate the
unknown constant $G$.

We proceed to briefly describe our calculation of the desired NRQCD
short-distance coefficients. The technical details will be presented
in a long write-up. Rather than invoking the standard perturbative
matching approach, we resort to a shortcut by directly extracting
the hard region contribution of a loop diagram in the spirit of
method of region~\cite{Beneke:1997zp}.

For the process at hand, the NRQCD factorization also holds at the
amplitude level. We start by considering the on-shell amplitude for
$c \bar{c}\to 3\gamma$ through ${\cal O}(\alpha_s)$.  At the tree
level, there are 6 diagrams for this process. At next-to-leading
order in $\alpha_s$, there are 12 self-energy diagrams, 12 outer
vertex diagrams, 6 inner vertex diagrams, 12 double vertex diagrams,
together with 6 ladder diagrams. The \textsc{Mathematica} package
\textsc{FeynArts}~\cite{Kublbeck:1990xc} is utilized to generate the
Feynman diagrams and the corresponding amplitude.

We use Dimensional Regularization to regularize both UV and IR
divergences. It is convenient to employ the covariant spin
projection technique~\cite{Bodwin:2002hg} to enforce the $c\bar{c}$
pair to be in the color-singlet and spin-triplet state. The color
and Dirac trace algebra are handled by the package
\textsc{FeynCalc}~\cite{Mertig:1990an}. Prior to carrying out the
loop integration, we expand the amplitude in powers of relative
momentum between $c$ and $\bar{c}$ through the second order,
followed by the projection of the $S$-wave orbital angular momentum
state.

We then employ two different versions of the self-written
\textsc{Mathematica} codes~\cite{Feng:2012iq} to carry out the
partial fractions, to reduce the general higher-point tensor
one-loop integrals into a minimal set of scalar integrals up to the
four-point. Up to this step, some scalar integrals may contain
propagator of cubic power.  The package
\textsc{FIRE}~\cite{Smirnov:2008iw} is then employed to reduce these
unconventional scalar integrals to the standard ones.  All the
required one-loop master integrals can be found in
Ref.~\cite{Adkins:2005eg}.

After summing the contributions from all the diagrams, and taking
into account the wave function renormalization of the charm quark,
the ultimate QCD amplitude becomes completely UV finite,
nevertheless contains a piece of unremoved logarithmic IR divergence
at the relative order $\alpha_s v^2$. The occurrence of the IR
divergence in the hard region at this order is just as expected,
physically because of the breakdown of the color transparency once
beyond the leading order in $v$. This IR divergence can be absorbed
in the $\overline{\rm MS}$-renormalized NRQCD matrix element, and
consequently, the ${\cal O}(\alpha_s v^2)$ short-distance
coefficient now explicitly depends on the NRQCD factorization scale
$\mu_f$, with a natural range from $m_c v$ to $m_c$.

Since the differential short-distance coefficients inferred at the
amplitude level are IR finite, one can safely return to 4 spacetime
dimensions. Squaring the amplitude, summing over the photon
polarizations and averaging upon the $J/\psi$ spin, integrating over
the entire three-photon phase space and multiplying a symmetry
factor ${1\over 3!}$, we are able to determine each of the
short-distance coefficients tabulated in
(\ref{NRQCD:factorization:decay:rate}). We readily confirm the
existing ${\cal O}(\alpha_s)$ and ${\cal O}(v^2)$
coefficients~\cite{Kwong:1987ak,Schuler:1994hy}.

We use the built-in function of multi-dimensional global adaptive
integration algorithm in \textsc{Mathematica} to compute the
one-loop coefficients. We finally nail down the value of the desired
constant to be
%-----------------------------
\bqa
%-----------------------------
G &=& 68.913,
%-----------------------------
\label{alphas:v2:coefficient}
%-----------------------------
\eqa
%-----------------------------
with the fractional error estimated to be smaller than $10^{-4}$.

Since the logarithmic term in the ${\cal O}(\alpha_s v^2)$
coefficient is overwhelmed by the $G$, from now on we will set
$\mu_f=m_c$ in (\ref{NRQCD:factorization:decay:rate}). The full
NRQCD prediction is then proportional to
$1-4.02\,\alpha_s(\mu)+[-5.32 + 21.94 \,\alpha_s(\mu)]\langle
v^2\rangle_{J/\psi}$. It is clear to see that the new piece of
correction yields a surprisingly significant and positive
contribution. With a reasonable choice of $\mu$, it has strong
potential to counterbalance the known large negative corrections.

In the numerical analysis, we take the values of the NRQCD matrix
elements from Ref.~\cite{Bodwin:2007fz}:
%-----------------------------
\bqa
%-----------------------------
\left|\langle 0| \chi^\dagger\bm{\sigma}\cdot
\bm{\epsilon}^\ast\psi|J/\psi(\bfepsilon)\rangle \right|^2
=0.446\:{\rm GeV}^3,\qquad \langle v^2\rangle_{J/\psi}=0.223,
%-----------------------------
\label{NRQCD:ME:values}
%-----------------------------
\eqa
%-----------------------------
which are fitted through the decay $J/\psi\to e^+e^-$ accurate
through the relative order $v^2$. We take the charm quark pole mass
as $m_c=1.4$ GeV. The Gremm-Kapustin relation~\cite{Gremm:1997dq}
then implies that the value of $\langle v^2\rangle_{J/\psi}$ is
consistent with that given in (\ref{NRQCD:ME:values}).

We take the fine structure constant $\alpha$ to be $1/137$. There
exists some ambiguity in choosing the strong coupling constant
$\alpha_s(\mu)$. Presumably, one may think that any scale between $2
m_c/3$ and $2m_c$ GeV could be equally acceptable.

Taking $\mu=m_c$ in (\ref{NRQCD:factorization:decay:rate}),
consequently $\alpha_s(m_c)=0.388$, which is calculated through the
two-loop renormalization group equation with $\Lambda^{(n_f=3)}_{\rm
QCD}=390$ MeV, we then obtain the NRQCD predictions for the decay
rate of $J/\psi\to 3\gamma$ at various level of accuracy. The LO
prediction is 6.01 eV. After including ${\cal O}(\alpha_s)$
correction, the prediction drops to $-3.36$ eV. If further including
the ${\cal O}(v^2)$ correction, we then get $-10.48$ eV. However,
once the new ${\cal O}(\alpha_s v^2)$ correction is added, we end up
with the reasonable value of 0.91 eV. Dividing these predicted
partial widths by the total width of $J/\psi$, whose latest value is
$92.9$ keV~\cite{Beringer:1900zz}, we then find the ${\rm
Br}[J/\psi\to 3\gamma]$ to be $6.46\times 10^{-5}$, $-3.61\times
10^{-5}$, $-11.28 \times 10^{-5}$, and $0.98\times 10^{-5}$,
respectively. It is amazing that after incorporating the new ${\cal
O}(\alpha_s v^2)$ correction, the full NRQCD prediction agrees with
both the \textsf{CLEO-c}~\cite{Adams:2008aa} and
\textsf{BESIII}~\cite{Ablikim:2012sn}
 data quite well.

One might be curious about the sensitivity of our predictions to the
renormalization scale $\mu$. To address this question, in
Fig.~\ref{Various:predictions:depend:mu} we explicitly illustrate
the scale dependence of various NRQCD predictions for ${\rm
Br}[J/\psi\to 3\gamma]$ in the range between 1 and 3 GeV. For
reader's convenience, the \textsf{BESIII}
measurement~\cite{Ablikim:2012sn} is also juxtaposed in the plot.
One readily observes that the full NRQCD prediction exhibits much
flatter $\mu$-dependence than that only incorporating the ${\cal
O}(\alpha_s)$ correction. Finally, it is interesting to remark that,
the agreement between the full NRQCD prediction and the data appears
to be rather stable in a window of between $1.1 \leq\mu \leq 1.5$
GeV.

In summary, in this work we investigate the ${\cal O}(\alpha_s v^2)$
correction to the rare decay process $J/\psi\to 3\gamma$ in the
NRQCD framework. Thus far, the large negative first-order radiative
and relativistic corrections have prevented one from making any
sensible prediction for the corresponding decay branching fraction.
It is very encouraging that this new ingredient of correction turns
out to be significant and positive. Including this ${\cal
O}(\alpha_s v^2)$ correction appears to be crucial to reconcile the
NRQCD prediction and the recent \textsf{CLEO-c} and \textsf{BESIII}
experiments. The satisfactory agreement between the data and theory
may lend some support to the values of the NRQCD matrix elements
obtained in Ref.~\cite{Bodwin:2007fz}. Finally, we remark that the
future exploration of the ${\cal O}(v^4)$ and ${\cal O}(\alpha_s^2)$
corrections for this process, may turn out to be quite enlightening.

%--------------------------------------------------------------------
\begin{acknowledgments}
%--------------------------------------------------------------------
This research was supported in part by the National Natural Science
Foundation of China under Grant No.~10875130, No.~10935012, DFG and
NSFC (CRC 110), and China Postdoctoral Science Foundation.
%--------------------------------------------------------------------
\end{acknowledgments}
%--------------------------------------------------------------------

%%%%%%%%%%%%%%%%%%%%%%%%%%%%%%%%%%%%%%%%%%%%%%%%%%%%%%%%%%%%%%%%%%%%%%%%%%%%%%


\begin{thebibliography}{150}

%\cite{Caswell:1985ui}
\bibitem{Caswell:1985ui}
  W.~E.~Caswell and G.~P.~Lepage,
  %``Effective Lagrangians for Bound State Problems in QED, QCD, and Other Field Theories,''
  Phys.\ Lett.\ B {\bf 167}, 437 (1986).  %%CITATION = PHLTA,B167,437;%%

%\cite{Caswell:1976nx}
\bibitem{Caswell:1976nx}
  W.~E.~Caswell, G.~P.~Lepage and J.~R.~Sapirstein,
  %``Order (alpha) Corrections to the Decay Rate of Orthopositronium,''
  Phys.\ Rev.\ Lett.\  {\bf 38} (1977) 488;  %%CITATION = PRLTA,38,488;%%

%\cite{Adkins:2005eg}
\bibitem{Adkins:2005eg}
  G.~S.~Adkins,
  %``Analytic evaluation of the amplitudes for orthopositronium decay to three photons to one-loop order,''
   Phys.\ Rev.\ Lett.\  {\bf 76}, 4903 (1996)  [hep-ph/0506213].  %%CITATION = HEP-PH/0506213;%%

\bibitem{Adkins:2000fg}
%\cite{Adkins:2000fg}
%\bibitem{Adkins:2000fg}
  G.~S.~Adkins, R.~N.~Fell and J.~R.~Sapirstein,
  %``Order alpha**2 corrections to the decay rate of orthopositronium,''
   Phys.\ Rev.\ Lett.\  {\bf 84}, 5086 (2000)  [hep-ph/0003028];
   %%CITATION = HEP-PH/0003028;%%
G. S. Adkins, R. N. Fell, and J. Sapirstein, Ann. Phys. (N.Y.) {\bf
295}, 136 (2002);
\\
%\cite{Hill:2000qi}
%\bibitem{Hill:2000qi}
  R.~J.~Hill and G.~P.~Lepage,
  %``Order (alpha**2 Gamma) binding effects in orthopositronium decay,''
  Phys.\ Rev.\ D {\bf 62}, 111301 (2000)  [hep-ph/0003277];  %%CITATION = HEP-PH/0003277;%%
\\
%\cite{Kniehl:2000dh}
%\bibitem{Kniehl:2000dh}
  B.~A.~Kniehl and A.~A.~Penin,
  %``Order alpha**3 ln (1 / alpha) corrections to positronium decays,''
  Phys.\ Rev.\ Lett.\  {\bf 85}, 1210 (2000)  [Erratum-ibid.\  {\bf 85}, 3065 (2000)]  [hep-ph/0004267];  %%CITATION = HEP-PH/0004267;%%
\\
%\cite{Melnikov:2000fi}
%\bibitem{Melnikov:2000fi}
  K.~Melnikov and A.~Yelkhovsky,
  %``O(alpha**3 ln alpha) corrections to positronium decay rates,''
  Phys.\ Rev.\ D {\bf 62}, 116003 (2000)  [hep-ph/0008099].  %%CITATION = HEP-PH/0008099;%%

%\cite{Jinnouchi:2003hr}
\bibitem{Jinnouchi:2003hr}
  O.~Jinnouchi, S.~Asai and T.~Kobayashi,
  %``Precision measurement of orthopositronium decay rate using SiO(2) powder,''
  Phys.\ Lett.\ B {\bf 572}, 117 (2003)  [hep-ex/0308030];  %%CITATION = HEP-EX/0308030;%%
%\cite{Vallery:2003iz}
%\bibitem{Vallery:2003iz}
  R.~S.~Vallery, P.~W.~Zitzewitz and D.~W.~Gidley,
  %``Resolution of the Orthopositronium Lifetime Puzzle,''
  Phys.\ Rev.\ Lett.\  {\bf 90}, 203402 (2003).  %%CITATION = PRLTA,90,203402;%%


%\cite{Mackenzie:1981sf}
%\bibitem{Mackenzie:1981sf}
%  P.~B.~Mackenzie and G.~P.~Lepage,
%  %``QCD Corrections to the Gluonic Width of the Upsilon Meson,''
%  Phys.\ Rev.\ Lett.\  {\bf 47}, 1244 (1981).  %%CITATION = PRLTA,47,1244;%%

%\cite{Adams:2008aa}
\bibitem{Adams:2008aa}
  G.~S.~Adams {\it et al.}  [CLEO Collaboration],
  %``Observation of J/psi --> 3 gamma,''
  Phys.\ Rev.\ Lett.\  {\bf 101}, 101801 (2008)  [arXiv:0806.0671 [hep-ex]].
  %%CITATION = ARXIV:0806.0671;%%

%\cite{Ablikim:2012sn}
\bibitem{Ablikim:2012sn}
  M.~Ablikim {\it et al.}  [BESIII Collaboration],
  %``Evidence for $\eta_{c} \rightarrow \gamma\gamma$ and Measurement of $J/\psi\rightarrow 3\gamma$,''
    arXiv:1208.1461 [hep-ex].  %%CITATION = ARXIV:1208.1461;%%


%\cite{Mackenzie:1981sf}
\bibitem{Mackenzie:1981sf}
  P.~B.~Mackenzie and G.~P.~Lepage,
  %``QCD Corrections to the Gluonic Width of the Upsilon Meson,''
  Phys.\ Rev.\ Lett.\  {\bf 47}, 1244 (1981).  %%CITATION = PRLTA,47,1244;%%

%\cite{Keung:1982jb}
\bibitem{Keung:1982jb}
  W.~-Y.~Keung and I.~J.~Muzinich,
  %``Beyond The Static Limit For Quarkonium Decays,''
  Phys.\ Rev.\ D {\bf 27}, 1518 (1983).
  %%CITATION = PHRVA,D27,1518;%%

%\cite{Kwong:1987ak}
\bibitem{Kwong:1987ak}
  W.~Kwong, P.~B.~Mackenzie, R.~Rosenfeld and J.~L.~Rosner,
  %``Quarkonium Annihilation Rates,''
  Phys.\ Rev.\ D {\bf 37}, 3210 (1988).
  %%CITATION = PHRVA,D37,3210;%%

%\cite{Brambilla:2010cs}
\bibitem{Brambilla:2010cs}
  N.~Brambilla {\it et al.},
  %S.~Eidelman, B.~K.~Heltsley, R.~Vogt, G.~T.~Bodwin,
  %E.~Eichten, A.~D.~Frawley and A.~B.~Meyer {\it et al.},
  %``Heavy quarkonium: progress, puzzles, and opportunities,''
  Eur.\ Phys.\ J.\ C {\bf 71}, 1534 (2011)  [arXiv:1010.5827 [hep-ph]].
  %%CITATION = ARXIV:1010.5827;%%

%\cite{Bodwin:1994jh}
\bibitem{Bodwin:1994jh}
  G.~T.~Bodwin, E.~Braaten and G.~P.~Lepage,
  %``Rigorous QCD analysis of inclusive annihilation and production of heavy
  %quarkonium,''
  Phys.\ Rev.\  D {\bf 51}, 1125 (1995)
  [Erratum-ibid.\  D {\bf 55}, 5853 (1997)]
  [arXiv:hep-ph/9407339].
  %%CITATION = PHRVA,D51,1125;%%

%\cite{Gremm:1997dq}
\bibitem{Gremm:1997dq}
  M.~Gremm and A.~Kapustin,
  %``Annihilation of S wave quarkonia and the measurement of alpha-s,''
  Phys.\ Lett.\ B {\bf 407}, 323 (1997)  [hep-ph/9701353].  %%CITATION = HEP-PH/9701353;%%

%\cite{Ore:1949te}
\bibitem{Ore:1949te}
  A.~Ore and J.~L.~Powell,
  %``Three photon annihilation of an electron - positron pair,''
  Phys.\ Rev.\  {\bf 75}, 1696 (1949).
  %%CITATION = PHRVA,75,1696;%%

%\cite{Schuler:1994hy}
\bibitem{Schuler:1994hy}
  G.~A.~Schuler,
  %``Quarkonium production and decays,''  %Submitted to: Phys.Rept.
  hep-ph/9403387.  %%CITATION = HEP-PH/9403387;%%

%\cite{Braaten:1998au}
\bibitem{Braaten:1998au}
  E.~Braaten and Y.~-Q.~Chen,
  %``Renormalons in electromagnetic annihilation decays of quarkonium,''
  Phys.\ Rev.\ D {\bf 57}, 4236 (1998)
   [Erratum-ibid.\ D {\bf 59}, 079901 (1999)]  [hep-ph/9710357].
   %%CITATION = HEP-PH/9710357;%%

%\cite{Luke:1997ys}
\bibitem{Luke:1997ys}
  M.~E.~Luke and M.~J.~Savage,
  %``Power counting in dimensionally regularized NRQCD,''
  Phys.\ Rev.\ D {\bf 57}, 413 (1998)  [hep-ph/9707313].  %%CITATION = HEP-PH/9707313;%%

%\cite{Bodwin:2008vp}
\bibitem{Bodwin:2008vp}
  G.~T.~Bodwin, H.~S.~Chung, J.~Lee and C.~Yu,
  %``Order-alpha(s) corrections to the quarkonium electromagnetic current at all orders in the heavy-quark velocity,''
  Phys.\ Rev.\ D {\bf 79}, 014007 (2009)  [arXiv:0807.2634 [hep-ph]].  %%CITATION = ARXIV:0807.2634;%%

%\cite{Lee:2010ts}
\bibitem{Lee:2010ts}
  J.~Lee, W.~Sang and S.~Kim,
  %``Relativistic corrections to the axial vector and vector currents in the bar{b}c meson system at order alpha_s,''
  JHEP {\bf 1101}, 113 (2011)  [arXiv:1011.2274 [hep-ph]].
  %%CITATION = ARXIV:1011.2274;%%


%\cite{Jia:2011ah}
\bibitem{Jia:2011ah}
  Y.~Jia, X.~-T.~Yang, W.~-L.~Sang and J.~Xu,
  %``$O(\alpha_s v^2)$ correction to pseudoscalar quarkonium decay to two photons,''
  JHEP {\bf 1106}, 097 (2011)  [arXiv:1104.1418 [hep-ph]].  %%CITATION = ARXIV:1104.1418;%%

%\cite{Guo:2011tz}
\bibitem{Guo:2011tz}
  H.~-K.~Guo, Y.~-Q.~Ma and K.~-T.~Chao,
  %``$O(\alpha_sv^2)$ Corrections to Hadronic and Electromagnetic Decays of $^1S_0$ Heavy Quarkonium,''
  Phys.\ Rev.\ D {\bf 83}, 114038 (2011)  [arXiv:1104.3138 [hep-ph]].  %%CITATION = ARXIV:1104.3138;%%

%\cite{Li:2012rn}
\bibitem{Li:2012rn}
  J.~-Z.~Li, Y.~-Q.~Ma and K.~-T.~Chao,
  %``QCD and Relativistic $O(\alpha_{s}v^2)$ Corrections to Hadronic Decays
  %of Spin-Singlet Heavy Quarkonia $h_c, h_b$ and $\eta_b$,''
  arXiv:1209.4011 [hep-ph].  %%CITATION = ARXIV:1209.4011;%%


%\cite{Dong:2012xx}
\bibitem{Dong:2012xx}
  H.~-R.~Dong, F.~Feng and Y.~Jia,
  %``$O(\alpha_s v^2)$ correction to $e^+e^-\to J/\psi+\eta_c$ at $B$ factories,''
  Phys.\ Rev.\ D {\bf 85}, 114018 (2012)  [arXiv:1204.4128 [hep-ph]].  %%CITATION = ARXIV:1204.4128;%%

%\cite{Beneke:1997zp}
\bibitem{Beneke:1997zp}
  M.~Beneke and V.~A.~Smirnov,
  %``Asymptotic expansion of Feynman integrals near threshold,''
  Nucl.\ Phys.\ B {\bf 522}, 321 (1998)  [hep-ph/9711391].  %%CITATION = HEP-PH/9711391;%%

%\cite{Kublbeck:1990xc}
\bibitem{Kublbeck:1990xc}
  J.~Kublbeck, M.~Bohm and A.~Denner,
  %``Feyn Arts: Computer Algebraic Generation Of Feynman Graphs And Amplitudes,''
  Comput.\ Phys.\ Commun.\  {\bf 60}, 165 (1990);  %%CITATION = CPHCB,60,165;%%
  \\
%\cite{Hahn:2000kx}
%\bibitem{Hahn:2000kx}
  T.~Hahn,
  %``Generating Feynman diagrams and amplitudes with FeynArts 3,''
  Comput.\ Phys.\ Commun.\  {\bf 140}, 418 (2001)  [hep-ph/0012260].  %%CITATION = HEP-PH/0012260;%%

%\cite{Bodwin:2002hg}
\bibitem{Bodwin:2002hg}
  G.~T.~Bodwin and A.~Petrelli,
  %``Order v**4 corrections to S wave quarkonium decay,''
  Phys.\ Rev.\ D {\bf 66}, 094011 (2002)  [hep-ph/0205210].  %%CITATION = HEP-PH/0205210;%%

%\cite{Mertig:1990an}
\bibitem{Mertig:1990an}
  R.~Mertig, M.~Bohm and A.~Denner,
  %``FEYN CALC: Computer algebraic calculation of Feynman amplitudes,''
  Comput.\ Phys.\ Commun.\  {\bf 64}, 345 (1991).  %%CITATION = CPHCB,64,345;%%

%\cite{Feng:2012iq}
\bibitem{Feng:2012iq}
  F.~Feng,
  %``$Apart: A Generalized Mathematica Apart Function,''
  Comput.\ Phys.\ Commun.\  {\bf 183}, 2158 (2012)  [arXiv:1204.2314 [hep-ph]].  %%CITATION = ARXIV:1204.2314;%%

%\cite{Smirnov:2008iw}
\bibitem{Smirnov:2008iw}
  A.~V.~Smirnov,
  %``Algorithm FIRE -- Feynman Integral REduction,''
  JHEP {\bf 0810}, 107 (2008)  [arXiv:0807.3243 [hep-ph]].
  %%CITATION = ARXIV:0807.3243;%%

%\cite{Bodwin:2007fz}
\bibitem{Bodwin:2007fz}
  G.~T.~Bodwin, H.~S.~Chung, D.~Kang, J.~Lee and C.~Yu,
  %``Improved determination of color-singlet nonrelativistic QCD matrix elements for S-wave charmonium,''
  Phys.\ Rev.\ D {\bf 77}, 094017 (2008)  [arXiv:0710.0994 [hep-ph]].  %%CITATION = ARXIV:0710.0994;%%

%\cite{Beringer:1900zz}
\bibitem{Beringer:1900zz}
  J.~Beringer {\it et al.}  [Particle Data Group Collaboration],
  %``Review of Particle Physics (RPP),''
  Phys.\ Rev.\ D {\bf 86}, 010001 (2012).  %%CITATION = PHRVA,D86,010001;%%


\end{thebibliography}
\end{document}